\documentclass[cameraready]{Interspeech}
\usepackage{multirow}

\title{Not All Frames Are Equal: \\ Difference-Aware Quantization for Ultra-Low-Bit ASR}

\author[affiliation={1}]{Woori}{Jeon}
\author[affiliation={1}, correspondingauthor]{Jungmin}{So}

\address{
    $^1$ Sogang University, Seoul, Korea
}

\email{woori50300@gmail.com, jso1@sogang.ac.kr}


\begin{document}
\maketitle

\begin{abstract}

Post-training quantization (PTQ) methods such as GPTQ and AWQ compress large language models effectively, but applying them to automatic speech recognition (ASR) at ultra-low bit-widths (2-3 bits) causes severe transcription degradation. Unlike discrete text tokens, speech activations are highly correlated in steady-state and zero-padded regions yet change rapidly at phonetic boundaries; when all frames contribute equally to the Hessian, static frames dominate calibration and obscure informative transitions. We propose DiffAQ, which computes frame-to-frame activation differences to measure the rate of acoustic change and assigns Hessian importance proportionally, concentrating quantization precision on phonetically critical frames. As a training-free modification to GPTQ, DiffAQ consistently reduces WER across various Whisper sizes and standard benchmarks, with the largest gains at 2-bit precision where baseline methods frequently produce degenerate outputs.

\end{abstract}

\section{Introduction}

The deployment of state-of-the-art Automatic Speech Recognition (ASR) foundation models, such as OpenAI's Whisper \cite{radford2023whisper}, on edge devices is heavily bottlenecked by memory footprint and bandwidth constraints. While these models achieve near-human transcription accuracy, their parameter counts preclude real-time execution on resource-constrained hardware. 

Recent advancements in Post-Training Quantization (PTQ), notably GPTQ \cite{frantar2022gptq} and AWQ \cite{lin2023awq}, have successfully mitigated this issue for Large Language Models (LLMs). These methods compress weights to extremely low precision in a matter of minutes without requiring the massive computational overhead of Quantization-Aware Training (QAT). GPTQ achieves this by using a second-order Hessian inverse matrix to distribute quantization errors optimally, while AWQ scales salient weights based on activation magnitudes.

\begin{figure}[t]
  \centering
  \includegraphics[width=\linewidth]{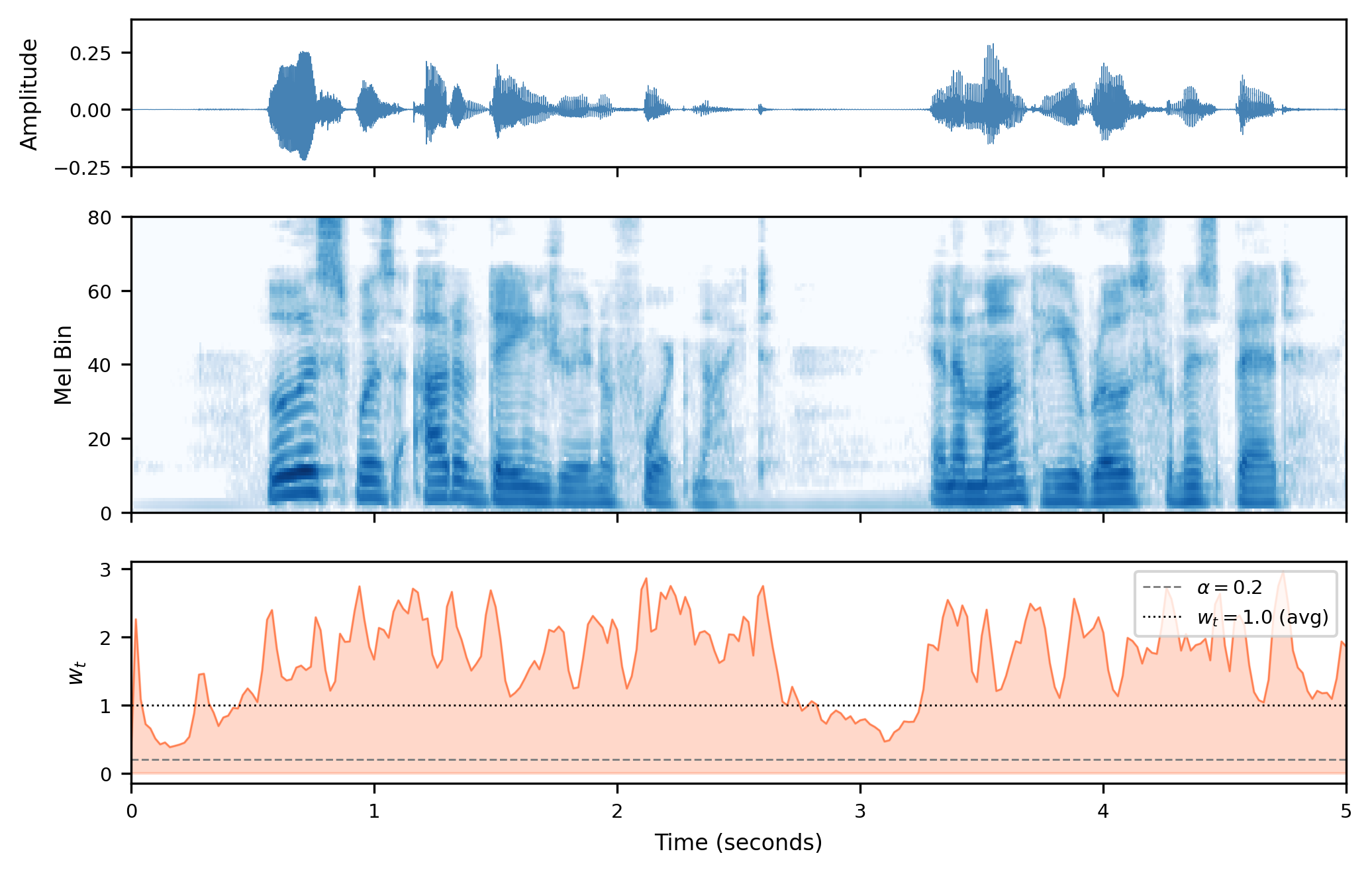}
  \caption{Waveform (top), Mel-spectrogram (middle), and DiffAQ importance weight $w_t$ (bottom) for a LibriSpeech utterance. $w_t$ drops toward the floor $\alpha{=}0.2$ during silence and rises above the mean ($w_t{=}1.0$) at phonetic transitions, giving acoustically dynamic frames greater importance during quantization calibration.}
  \label{fig:wt_overlay}
\end{figure}

However, researchers attempting to directly apply standard GPTQ and AWQ to ASR models have observed severe transcription degradation often resulting in complete hallucination loops at 2-bit and 3-bit weight precisions \cite{feng2025edgeasr}. We attribute this failure to a fundamental \textit{modality gap} between text and speech \cite{baevski2020wav2vec}. Text consists of discrete tokens with uniform information density. In contrast, ASR models process continuous audio representations that are heavily padded, temporally correlated, and acoustically redundant, as the waveform and spectrogram in Figure~\ref{fig:wt_overlay} show. Standard PTQ evaluates parameter importance by treating every temporal frame's magnitude equally.

To bridge this modality gap, we introduce \textbf{Difference-Aware Quantization (DiffAQ)}. Since speech intelligibility relies predominantly on dynamic spectral transitions \cite{stevens2002toward, greenberg1999speaking}, we modify the PTQ calibration process to track frame-to-frame activation differences. DiffAQ assigns Hessian importance proportional to the rate of acoustic change, concentrating quantization precision on frames where phonetically distinctive information is concentrated. Figure~\ref{fig:wt_overlay} visualizes this: $w_t$ drops during silence and rises sharply at phonetic boundaries.

Our core contributions are as follows:
\begin{itemize}
    \item We identify that standard PTQ methods are ill-suited for ASR at ultra-low bit-widths (2-3 bits): their frame-agnostic Hessian estimation is dominated by temporally redundant and zero-padded frames, leaving the model poorly calibrated for acoustic transitions that drive transcription accuracy.
    \item We propose DiffAQ, which computes frame-to-frame activation differences to measure the rate of acoustic change and incorporates this as an importance weight during Hessian accumulation, ensuring that quantization error is minimized at acoustically dynamic frames where phonetically distinctive information is concentrated.
    \item We validate DiffAQ across three Whisper sizes (base, small, and medium), demonstrating consistent WER improvements at 2-bit and 3-bit quantization over standard GPTQ and AWQ baselines. As a training-free modification to GPTQ, DiffAQ yields the largest gains at 2-bit precision where baseline methods frequently produce degenerate outputs.
\end{itemize}

\section{Background and Related Work}
\label{sec:background}
\subsection{Activation-Aware Post-Training Quantization}
The most straightforward PTQ approach is Round-to-Nearest (RTN), which quantizes each weight by rounding it to the nearest representable value at the target bit-width. While computationally efficient, RTN ignores the relative importance of individual weights, leading to severe accuracy degradation at ultra-low bit-widths  \cite{jacob2018quantization}.
Modern PTQ relies on calibration data to identify which weights are most important. GPTQ, rooted in Optimal Brain Quantization (OBQ) \cite{frantar2022gptq}, minimizes the squared output error using the empirical Hessian matrix of the activations:
\begin{equation}
  \mathbf{H} = 2 \mathbf{X} \mathbf{X}^\top = 2 \sum_{t=1}^{T} \mathbf{x}_t \mathbf{x}_t^\top
  \label{eq:standard_hessian}
\end{equation}
AWQ \cite{lin2023awq} takes a different approach, protecting salient weights by observing the raw magnitude of the activation channels ($\mathbf{s} = \frac{1}{T} \sum |\mathbf{x}_t|$) and mathematically scaling the corresponding weights before RTN quantization.

\subsection{ASR Model Quantization}
Model compression for ASR has relied on knowledge distillation \cite{gandhi2024distilwhisper} or QAT \cite{rybakov20232bit, shao2024dqwhisper} to reach low bit-widths. While effective, these approaches require significant computational resources and access to transcribed training data.

PTQ avoids retraining entirely, but early attempts such as Q-ASR \cite{kim2022integer} reached only 8-bit precision, and ultra-low-bit quantization remains challenging. Fish et al. \cite{fish2023model} proposed myQASR, a mixed-precision PTQ method based on layer sensitivity. However, these methods, like standard GPTQ and AWQ, still calibrate using raw activation magnitudes without accounting for the temporal structure of speech.

\subsection{Challenges of PTQ for Speech}
Unlike LLMs, ASR models operate on log-Mel spectrograms where a \SI{25}{\milli\second} window advances in \SI{10}{\milli\second} steps, creating 60\% overlap between adjacent frames. Due to this overlap, steady-state segments such as sustained vowels, silence, and padding produce nearly static consecutive frames \cite{baevski2020wav2vec, hsu2021hubert}, and architectures such as Squeezeformer \cite{kim2022squeezeformer} and FastConformer \cite{rekesh2023fast} explicitly downsample the temporal axis to address this redundancy.

When all frames contribute equally to the Hessian (Eq.~\ref{eq:standard_hessian}), these static frames disproportionately dominate the importance estimate, leaving the quantized model poorly calibrated for the acoustic transitions that drive transcription accuracy. This problem is amplified in Whisper, which forces all inputs into a fixed 30-second context window (3,000 frames). If a calibration utterance contains only 3 seconds of active speech, the remaining 2,700 frames consist of zero-padding \cite{bain2023whisperx}. GPTQ's Cholesky solver then optimizes the quantized weights to reconstruct 27 seconds of silence, largely exhausting the network's capacity to preserve the brief transients that carry phonetically distinctive information.

\section{Difference-Aware Quantization (DiffAQ)}
\label{sec:method}

\begin{figure*}[t]
  \centering
  \includegraphics[width=\textwidth]{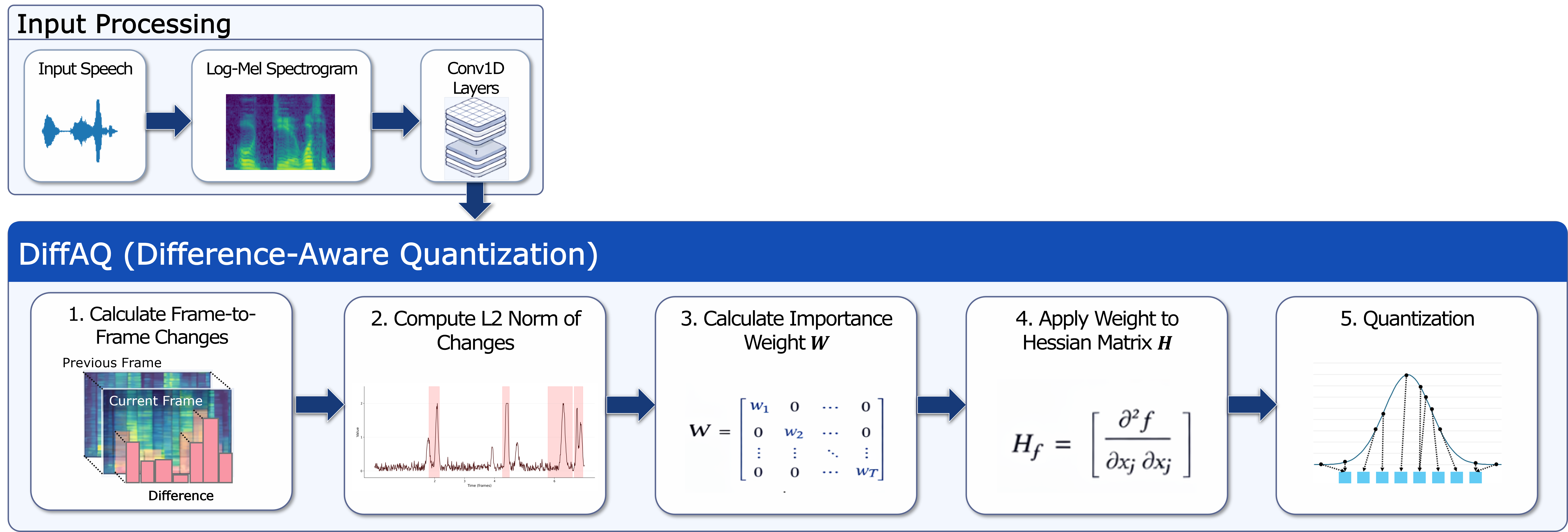}
  \caption{Overview of the DiffAQ encoder quantization pipeline. Input speech is converted to a log-Mel spectrogram and processed by Conv1D layers. DiffAQ then computes frame-to-frame activation differences, derives the importance weight $w_t$, and applies it to scale the Hessian before quantization.}
  \label{fig:pipeline}
\end{figure*}

Meaningful information in speech is concentrated in regions of rapid acoustic change \cite{stevens2002toward}, and spectral dynamics carry more discriminative power than static coefficients \cite{furui1986speaker}. We propose DiffAQ, which computes the frame-to-frame activation difference and incorporates it as an importance weight during Hessian accumulation. Although DiffAQ operates on intermediate hidden activations rather than raw audio, our experiments show that activation differences effectively distinguish dynamic from static regions.

\subsection{Frame-to-Frame Temporal Differencing}
Let $\mathbf{x}_t \in \mathbb{R}^D$ denote the $D$-dimensional hidden activation vector at frame $t$, extracted from an intermediate layer of the ASR encoder.

In classical speech feature extraction, dynamic coefficients are computed as temporal derivatives of the feature trajectory \cite{furui1986speaker}. For discrete frames, this is approximated via the finite difference:
\begin{equation}
  \frac{\partial \mathbf{x}}{\partial t} \approx \frac{\mathbf{x}_{t} - \mathbf{x}_{t-1}}{\Delta t}
\end{equation}
Because the frames are uniformly spaced, $\Delta t$ reduces to a constant and can be set to $1$.

This simplifies to a simple vector subtraction:
\begin{equation}
  \Delta \mathbf{x}_t = \mathbf{x}_{t} - \mathbf{x}_{t-1}
  \label{eq:diffing}
\end{equation}

We calculate the temporal density score, $d_t$, as the $L_2$ norm of this difference:

\begin{equation}
  d_t = \lVert \Delta \mathbf{x}_t \rVert_2
\end{equation}

This frame-to-frame differencing acts as a simple importance measure for the speech modality:
\begin{itemize}
    \item \textbf{Padding Eradication:} The difference between consecutive zero-padded frames is exactly zero, so padding receives minimal Hessian weight, a natural special case of the framework rather than a separate mechanism.
    \item \textbf{Redundant Frame Suppression:} Due to the \SI{15}{\milli\second} spectrogram overlap, temporally static segments (sustained vowels, background hum) produce highly correlated consecutive activations, yielding a near-zero difference and correspondingly low weight. This aligns with RASTA filtering \cite{hermansky1994rasta}, which suppresses slowly varying spectral components to improve robustness. Because these frames are already well represented by their neighbors, reducing their Hessian contribution does not discard information.
    \item \textbf{Spectral Transition Emphasis:} At phonetic boundaries (e.g., from a fricative to a vowel), the acoustic state changes rapidly. These transient frames yield a large $d_t$ and receive maximal Hessian weight, ensuring quantization error is minimized precisely where phonetically distinctive information is concentrated.
\end{itemize}

\subsection{Normalized Importance Weighting}
However, assigning zero importance to static frames can make the Hessian singular.

To prevent this, we normalize $d_t$ by its sequence mean $\mu_d = \frac{1}{T}\sum_t d_t$, yielding $\bar{d}_t = d_t / \mu_d$. We then define the temporal importance weight $w_t$ by introducing a base importance factor $\alpha$:
\begin{equation}
  w_t = \alpha + (1 - \alpha) \bar{d}_t
  \label{eq:importance_weight}
\end{equation}

Under this normalization, frames with average acoustic change receive $w_t = 1.0$, matching standard GPTQ weighting, while padding and steady-state frames are suppressed toward $\alpha$ and rapid transitions are amplified beyond unity. Here, $\alpha$ is a small constant that ensures no frame is fully discarded from the Hessian accumulation. We set $\alpha = 0.2$ throughout all experiments.

Finally, the DiffAQ Hessian matrix is accumulated by scaling the standard outer product of each frame's activation by its corresponding importance weight:
\begin{equation}
  \mathbf{H}_{\text{DiffAQ}} = 2 \sum_{t=1}^{T} w_t (\mathbf{x}_t \mathbf{x}_t^\top)
  \label{eq:diffq_hessian}
\end{equation}

By substituting this modified Hessian into the standard quantization solver, DiffAQ prioritizes dynamic acoustic transitions while maintaining the numerical stability required for low-bit compression.

\subsection{Quantization Pipeline}
Figure~\ref{fig:pipeline} illustrates the overall DiffAQ pipeline. Given a calibration set of audio inputs, we extract intermediate hidden activations from the encoder layers, compute the frame-to-frame difference $\Delta \mathbf{x}_t$, and derive the importance weight $w_t$ (Eq.~\ref{eq:importance_weight}) that scales each frame's contribution to the Hessian (Eq.~\ref{eq:diffq_hessian}). Because DiffAQ addresses temporal redundancy in continuous audio frames, a problem specific to the encoder, we apply it only to the encoder linear layers. The decoder autoregressively generates discrete text tokens and does not exhibit this redundancy \cite{peng2024owsm}, so we retain standard GPTQ (Eq.~\ref{eq:standard_hessian}) for the decoder.

\begin{table*}[t]
  \caption{Word Error Rate (WER \%) on LibriSpeech across different quantization methods for Whisper Medium, Small, and Base. Results are reported as mean ($\pm$95\% confidence interval) across calibration seeds. ($>100\%$ WER indicates degenerate output).}
  \label{tab:wer_results}
  \centering
  \begin{tabular}{l l c c c c c c}
    \toprule
    \multirow{2}{*}{\textbf{Method}} & \multirow{2}{*}{\textbf{Precision}} & \multicolumn{2}{c}{\textbf{Whisper Medium}} & \multicolumn{2}{c}{\textbf{Whisper Small}} & \multicolumn{2}{c}{\textbf{Whisper Base}} \\
    \cmidrule(lr){3-4} \cmidrule(lr){5-6} \cmidrule(lr){7-8}
    & & \textbf{clean} & \textbf{other} & \textbf{clean} & \textbf{other} & \textbf{clean} & \textbf{other} \\
    \midrule
    FP16 Baseline & 16-bit & 2.73 & 5.69 & 3.42 & 7.54 & 5.17 & 12.33 \\
    \midrule
    RTN & 2-bit & 40.38 $\pm$ 0.13 & 66.41 $\pm$ 0.18 & 99.32 $\pm$ 0.03 & $>100$ & $>100$ & $>100$ \\
    AWQ & 2-bit & 40.39 $\pm$ 0.13 & 66.27 $\pm$ 0.18 & 99.61 $\pm$ 0.02 & 99.60 $\pm$ 0.02 & $>100$ & $>100$ \\
    Standard GPTQ & 2-bit & 13.73 $\pm$ 1.67 & 17.53 $\pm$ 1.36 & 36.64 $\pm$ 2.10 & 51.30 $\pm$ 2.24 & $>100$ & $>100$ \\
    \textbf{DiffAQ (Ours)} & \textbf{2-bit} & \textbf{7.93 $\pm$ 1.70} & \textbf{12.93 $\pm$ 1.34} & \textbf{28.02 $\pm$ 2.35} & \textbf{39.26 $\pm$ 2.70} & \textbf{99.50 $\pm$ 0.02} & \textbf{$>100$} \\
    \midrule
    RTN & 3-bit & 2.86 $\pm$ 0.02 & 6.65 $\pm$ 0.03 & 4.87 $\pm$ 0.04 & 10.81 $\pm$ 0.03 & 13.66 $\pm$ 0.07 & 27.95 $\pm$ 0.08 \\
    AWQ & 3-bit & 2.86 $\pm$ 0.02 & 6.62 $\pm$ 0.03 & 4.87 $\pm$ 0.04 & 10.81 $\pm$ 0.03 & 13.55 $\pm$ 0.07 & 27.94 $\pm$ 0.08 \\
    Standard GPTQ & 3-bit & 2.90 $\pm$ 0.03 & 6.21 $\pm$ 0.05 & 7.11 $\pm$ 1.47 & 11.75 $\pm$ 1.42 & 9.81 $\pm$ 1.65 & 19.62 $\pm$ 1.41 \\
    \textbf{DiffAQ (Ours)} & \textbf{3-bit} & \textbf{2.81 $\pm$ 0.02} & \textbf{6.08 $\pm$ 0.05} & \textbf{4.64 $\pm$ 1.62} & \textbf{9.88 $\pm$ 1.90} & \textbf{8.45 $\pm$ 0.86} & \textbf{18.24 $\pm$ 0.71} \\
    \bottomrule
  \end{tabular}
\end{table*}

\begin{table*}[t]
  \caption{Word Error Rate (WER \%) on FLEURS across different quantization methods for Whisper Medium, Small, and Base. Results are reported as mean ($\pm$95\% confidence interval) across calibration seeds. ($>100\%$ WER indicates degenerate output).}
  \label{tab:wer_results_fleurs}
  \centering
  \begin{tabular}{l l c c c}
    \toprule
    \textbf{Method} & \textbf{Precision} & \textbf{Whisper Medium} & \textbf{Whisper Small} & \textbf{Whisper Base} \\
    \midrule
    FP16 Baseline & 16-bit & 4.58 & 6.33 & 9.85 \\
    \midrule
    RTN & 2-bit & 76.65 $\pm$ 0.49 & $>100$ & $>100$ \\
    AWQ & 2-bit & 75.80 $\pm$ 0.39 & 99.80 $\pm$ 0.00 & $>100$ \\
    Standard GPTQ & 2-bit & 18.05 $\pm$ 3.76 & 66.55 $\pm$ 4.99 & $>100$ \\
    \textbf{DiffAQ (Ours)} & \textbf{2-bit} & \textbf{12.07 $\pm$ 0.86} & \textbf{49.10 $\pm$ 4.29} & \textbf{$>100$} \\
    \midrule
    RTN & 3-bit & 5.01 $\pm$ 0.33 & 9.23 $\pm$ 0.01 & 31.51 $\pm$ 0.17 \\
    AWQ & 3-bit & 5.01 $\pm$ 0.33 & 9.23 $\pm$ 0.01 & 31.79 $\pm$ 0.19 \\
    Standard GPTQ & 3-bit & 5.08 $\pm$ 0.33 & 11.37 $\pm$ 3.01 & 21.23 $\pm$ 0.33 \\
    \textbf{DiffAQ (Ours)} & \textbf{3-bit} & \textbf{4.84 $\pm$ 0.07} & \textbf{8.69 $\pm$ 1.22} & \textbf{17.62 $\pm$ 2.30} \\
    \bottomrule
  \end{tabular}
\end{table*}

\section{Experimental Results}

\subsection{Datasets and Metrics}
We evaluate DiffAQ using three sizes of the Whisper architecture: \texttt{base.en} ($\sim$74M parameters), \texttt{small.en} ($\sim$244M parameters), and \texttt{medium.en} ($\sim$769M parameters). We report Word Error Rate (WER) on two benchmarks: LibriSpeech \cite{panayotov2015librispeech} (\texttt{test-clean} and \texttt{test-other}, where \texttt{test-other} contains more acoustically challenging conditions) and the English subset of FLEURS \cite{conneau2023fleurs} as an out-of-domain benchmark. For calibration, we randomly sample 128 utterances from the LibriSpeech \texttt{train-other-500} split. All results are averaged over 10 random calibration seeds and reported with 95\% confidence intervals.

\subsection{Baselines and Configuration}
We compare DiffAQ against three primary baselines:
\begin{enumerate}
    \item \textbf{Round-To-Nearest (RTN) \cite{jacob2018quantization}:} Naive uniform quantization without activation-aware compensation.
    \item \textbf{AWQ \cite{lin2023awq}:} Scales salient channels using activation magnitudes before RTN.
    \item \textbf{GPTQ \cite{frantar2022gptq}:} Minimizes layer-wise reconstruction error using the standard Hessian (Eq.~\ref{eq:standard_hessian}).
\end{enumerate}

All models are quantized asymmetrically using per-group quantization (group size 64). We target ultra-low bit-widths: 3-bit and 2-bit weight precision (W3A16 and W2A16). DiffAQ introduces only element-wise vector subtraction and normalization over the $T \times D$ activation matrix, which is asymptotically negligible compared to GPTQ's $O(D^3)$ Cholesky decomposition; in practice, calibration time is virtually identical to standard GPTQ.

\subsection{Ultra-Low Bit Transcription Accuracy}
Tables~\ref{tab:wer_results} and~\ref{tab:wer_results_fleurs} present the WER of DiffAQ and baselines under 3-bit and 2-bit quantization.

At 3-bit precision, DiffAQ achieves the lowest mean WER across all datasets and model sizes. For Whisper Medium, the gains over GPTQ are modest (e.g., 6.08\% vs.\ 6.21\% on \textit{test-other}), indicating that 3-bit precision retains sufficient capacity for standard methods. The gap widens for smaller models, which have less parameter redundancy and are therefore more sensitive to how calibration distributes quantization error; for Whisper Small on FLEURS, DiffAQ reduces WER from 11.37\% (GPTQ) to 8.69\%.

The advantage of DiffAQ becomes evident under the tighter constraints of 2-bit quantization. RTN and AWQ lack Hessian-based error compensation and suffer severe degradation at ultra-low bit-widths, often producing degenerate outputs that inflate the WER beyond 100\%. Notably, AWQ yields error rates nearly identical to RTN; we observe that at this precision, AWQ's top 1\% activation selection often identifies only a single channel group, providing insufficient compensation. While standard GPTQ partially mitigates this through Hessian-guided reconstruction, DiffAQ further reduces errors by concentrating Hessian weight on dynamic phonetic boundaries. For the Medium model, DiffAQ lowers the 2-bit WER from 17.53\% (GPTQ) to 12.93\% on the LibriSpeech \textit{test-other} split.

Finally, the Whisper Base model ($\sim$74M parameters) fails to produce meaningful transcriptions (WER $> 100\%$) at 2-bit precision under all tested methods, including DiffAQ. This suggests that the bottleneck is representational capacity rather than calibration strategy; improved weighting cannot compensate when the model simply lacks enough parameters to encode information at 2 bits per weight.

We also examine the sensitivity of DiffAQ to the floor parameter $\alpha$. Varying $\alpha \in \{0.0, 0.1, 0.2, 0.3\}$ for Whisper Small at 2-bit precision, the average WER differs by less than 0.2\% on LibriSpeech and 1\% on FLEURS across all values, confirming that the benefit of DiffAQ derives primarily from the temporal difference weighting rather than the specific choice of floor parameter. Even $\alpha = 0.0$ performs comparably, though a nonzero floor helps ensure numerical stability of the Hessian.

\section{Discussion}
\label{sec:discussions}
While DiffAQ shows consistent improvements, three limitations are worth noting. First, the temporal difference is not speech-selective; transient non-speech sounds such as background noise or sudden environmental events can produce large activation differences and receive unintended high importance during calibration. Incorporating a voice activity detection signal could improve robustness in noisy conditions. Second, for smaller models such as Whisper Base, combining DiffAQ with layer-wise mixed-precision allocation, where sensitive layers retain higher bit-widths, may offer a practical path forward when uniform 2-bit quantization exceeds the model's representational capacity. Third, our evaluation is limited to Whisper, which we chose as it is one of the most widely used open-source ASR models and whose fixed 30-second context window represents a worst case for the padding trap. Because the underlying temporal redundancy of mel-spectrogram features is shared by all encoder-based ASR models, we expect the principle to transfer to streaming and CTC-based \cite{graves2006connectionist} architectures, though empirical verification remains future work.

\section{Conclusions}
We proposed DiffAQ, a training-free modification to GPTQ that addresses the mismatch between standard PTQ calibration and speech inputs by weighting each frame's Hessian contribution according to its frame-to-frame activation difference. Across two benchmarks and three Whisper model sizes, DiffAQ consistently reduced WER at both 2-bit and 3-bit precision, with the largest gains at 2-bit: on Whisper Medium, DiffAQ lowered the \textit{test-other} WER from 17.53\% (GPTQ) to 12.93\%, while standard methods frequently produced degenerate outputs on smaller models. These results demonstrate that explicitly accounting for the temporal structure of speech during PTQ calibration is a practical step toward reliable ultra-low-bit ASR quantization. At 2-bit precision, the weight memory of Whisper Medium shrinks from approximately 1.5\,GB (FP16) to under 200\,MB, bringing practical ASR closer to mobile and embedded deployment. Extending DiffAQ to other encoder-based ASR architectures, combining it with mixed-precision layer allocation, and exploring joint weight-activation quantization are promising directions for future work.

\section{Use of Generative AI Disclosure}
Generative AI tools (Google Gemini, Anthropic Claude) were used for code development assistance and manuscript editing. All analyses, experimental design, and scientific conclusions were made by the authors.

\bibliographystyle{IEEEtran}
\bibliography{mybib}

\end{document}